\documentclass[%
 aip,
 amsmath,amssymb,
 reprint,%
]{revtex4-1}

\usepackage{graphicx}
\usepackage{dcolumn}
\usepackage{bm}
\usepackage[utf8]{inputenc}
\usepackage[T1]{fontenc}
\usepackage{mathptmx}
\usepackage{etoolbox}

\makeatletter
\def\@email#1#2{%
 \endgroup
 \patchcmd{\titleblock@produce}
  {\frontmatter@RRAPformat}
  {\frontmatter@RRAPformat{\produce@RRAP{*#1\href{mailto:#2}{#2}}}\frontmatter@RRAPformat}
  {}{}
}%
\makeatother
\begin{document}

\preprint{Ain-plane/123-QED}


\title[TEM analysis of \textit{a-}Tb$_{17}$Co$_{83}$]{Exploring the origins of perpendicular magnetic anisotropy in amorphous Tb-Co via changes in medium-range ordering}
\author{Ellis Kennedy}
\affiliation{Department of Materials Science and Engineering, University of California Berkeley, Berkeley, California, 94720, USA}
\author{Emily Hollingworth}
\affiliation{Department of Physics, University of California Berkeley, Berkeley, California, 94720, USA}
\author{Alejandro Ceballos}
\affiliation{Department of Materials Science and Engineering, University of California Berkeley, Berkeley, California, 94720, USA}
\author{Daisy O'Mahoney}
\affiliation{Department of Materials Science and Engineering, University of California Berkeley, Berkeley, California, 94720, USA}
\author{Colin Ophus}
\affiliation{NCEM, Molecular Foundry, Lawrence Berkeley National Laboratory, Berkeley, California, 94720, USA}
\author{Frances Hellman}
\affiliation{Department of Materials Science and Engineering, University of California Berkeley, Berkeley, California, 94720, USA}
\affiliation{Department of Physics, University of California Berkeley, Berkeley, California, 94720, USA}
\affiliation{Materials Science Division, Lawrence Berkeley National Laboratory, Berkeley, California, 94720, USA}
\author{M.C. Scott}
\email{mary.scott@berkeley.edu}
\affiliation{Department of Materials Science and Engineering, University of California Berkeley, Berkeley, California, 94720, USA}
\affiliation{NCEM, Molecular Foundry, Lawrence Berkeley National Laboratory, Berkeley, California, 94720, USA}
\date{\today}

\begin{abstract}
Amorphous thin films of Tb$_{17}$Co$_{83}$ (\textit{a}-Tb-Co) grown by magnetron co-sputtering exhibit changes in magnetic anisotropy with varying growth and annealing temperatures. The magnetic anisotropy constant increases with increasing growth temperature, which is reduced or vanishes upon annealing at temperatures above the growth temperature. The proposed explanation for this anisotropy in high orbital moment Tb-based transition metal alloys is an amorphous phase texturing with preferential in-plane and out-of-plane local bonding configurations for the rare-earth and transition metal atoms. Scanning nanodiffraction performed in a transmission electron microscope (TEM) is applied to \textit{a}-Tb$_{17}$Co$_{83}$ films deposited over a range of temperatures to measure relative changes in medium-range ordering (MRO). These measurements reveal an increase in MRO with higher growth temperatures and a decrease in MRO with higher annealing temperatures. The trend in MRO indicates a relationship between the magnetic anisotropy and local atomic ordering. Tilting select films the TEM measures variations in the local atomic structure a function of orientation within the films. The findings support claims that preferential ordering along the growth direction results from temperature-mediated adatom configurations during deposition, and that oriented MRO correlates with the larger anisotropy constants.

\end{abstract}

\maketitle

\section*{\label{sec:level1}Introduction}
Amorphous rare earth – transition metal (RE-TM) alloys have tunable magnetic properties, such as compensation temperature and perpendicular magnetic anisotropy (PMA) \cite{ding2013,andreenko1997,hasegawa1974, hellman1992,gambino1998,harris1994,harris1993}. These properties make them desirable as novel spintronic materials and for ultrafast magneto-optical recording devices. For REs with non-zero total orbital angular momentum, such as Tb, their magnetic anisotropy is believed to originate from the single ion anisotropy of the RE atoms as well as complex pair-bonding and preferential atomic configurations induced during film growth and subsequent thermal treatments \cite{hellman1992}. Out-of-plane (OOP) RE-TM bonding and in-plane (IP) TM-TM bonding have been correlated to bulk perpendicular magnetic anisotropy (PMA) in \textit{a-}RE-TM systems \cite{hellman1992,harris1993,harris1994}. The amorphous nature of these systems complicates their analysis, as variations in magnetic anisotropy must instead be attributed to subtle variations in local atomic ordering \cite{cochrane1978}. Previous work shows that PMA is independent of film thickness, surface layer magnetic interactions, and macroscopic growth-induced strain \cite{hellman1992}.

The succinct description of crystalline structures with defined unit cells and permitted symmetry operations does not extend to amorphous materials. Their lack of translational and rotational symmetry requires a statistical approach for analysis \cite{voyles2002,treacy2005}. Two-body and multi-body distribution functions are applied to the study of amorphous structures to determine the probability that two or more atoms will be separated by a specific distance. Short-range ordering (SRO) is probed through two-body statistical analysis, such as radial distribution functions. However, SRO is limited to the first coordination shell of the constituent atom. Looking out a little further, medium-range ordering (MRO) on the 1 – 5 nm length scale, can be probed with transmission electron microscopy (TEM) \cite{williamson1995,voyles2002}. Statistical analysis of MRO can determine the degree and type of ordering that exists between the extremes of long-range and short-range order \cite{nakhmanson2001,treacy2005}.

In this work, \textit{a-}Tb$_{17}$Co$_{83}$ is used as a representative \textit{a-}RE-TM system.  Tb (RE) atoms have more than half-filled \textit{4f}-electron orbitals and magnetic moments that align antiferromagnetically to the moments of the Co (TM) atoms \cite{hansen1991}. The Curie temperature (T$_C$), saturation moment, and compensation temperature (T$_{comp}$) are solely dependent on composition, while PMA and coercivity depend as well on growth temperature, and other deposition parameters \cite{hellman1992,hasegawa1974}. A series of \textit{a-}Tb$_{17}$Co$_{83}$ films were deposited at 20$^{\circ}$C, 200$^{\circ}$C, and 300$^{\circ}$C. The films then either received no further heat treatment or were annealed at 200$^{\circ}$C or 300$^{\circ}$C. IP and OOP magnetization vs magnetic field measurements were collected from the films. Congruent with previous studies, the PMA was found to increase with growth temperature and decrease with annealing temperature\cite{mergel1993,hasegawa1974,ceballos2021,hellman1992}. The TEM method of scanning nanodiffraction was used to probe the underlying structural mechanisms responsible for variations in the magnetic properties of \textit{a-}Tb$_{17}$Co$_{83}$ as a function of film deposition and annealing temperature. The relative MRO across the series of samples is measured with fluctuation electron microscopy (FEM), a specialized application of scanning nanodiffraction that is sensitive to changes in diffracted intensity related to variations in atomic configurations within an amorphous system \cite{voyles2002}. We determine that MRO in the films increases with higher deposition temperatures and decreases with higher annealing temperatures. The results support the model of magnetism in \textit{a-}RE-TM systems in which PMA is related to atom-specific preferential local atomic ordering.

\section*{\label{sec:level2}Experimental Details}

\textit{a-}Tb$_{17}$Co$_{83}$ films were produced using magnetron co-sputtering at 1.8 mTorr Ar pressure from separate Tb and Co targets. The base pressure of the chamber was 7 x 10$^{-8}$ Torr.  The \textit{a-}Tb$_{17}$Co$_{83}$ films were sputtered at room temperature (20$^{\circ}$C), 200$^{\circ}$C, and 300$^{\circ}$C with a capping layer of \textit{a-}SiN$_x$ sputtered at 3.0 mTorr of Ar pressure. The films were deposited on Norcada grids with 10 nm thick \textit{a-}SiN$_x$ membranes. Samples consisted of 30 nm \textit{a-}Tb$_{17}$Co$_{83}$ films capped with 10 nm of \textit{a-}SiN$_x$ to prevent oxidation. The 30 nm thickness was selected for the \textit{a-}Tb$_{17}$Co$_{83}$, as it was experimentally determined to produce a strong speckle intensity in scanning nanodiffraction. A control sample of 10 nm of \textit{a-}SiN$_x$ was sputtered at 3.0 mTorr of Ar pressure to determine the influence of the capping layer on the diffraction data. After deposition at 20$^{\circ}$C, some films were annealed at 200$^{\circ}$C or 300$^{\circ}$C for an hour under high vacuum. 

The magnetization as a function of applied magnetic field IP and OOP was measured in a Quantum Design Magnetic Properties Measurement System (MPMS) at 20$^{\circ}$C for a series of samples grown and annealed under identical conditions to those detailed above (Supplemental Materials). The films grown at 20$^{\circ}$C were annealed for an hour at 150$^{\circ}$C, 200$^{\circ}$C, 275$^{\circ}$C, and 350$^{\circ}$C to establish how magnetic anisotropy depends on annealing temperature. The applied field was varied between - 1 and 1 T, a range exceeding the magnetic saturation values for all samples. The intrinsic uniaxial anisotropy constant, K$_{ui}$, is calculated using the relation below, where M$_{s}$ is the saturation magnetization and H$_{k}$ is the anisotropy field required to fully magnetize the films IP.\cite{hellman1999,ceballos2021} 
\begin{eqnarray}
K_{ui} = \frac{H_{k}M_{s}}2 + 2\pi M_s^{2}
\end{eqnarray}

For all samples, Tc = 190$^{\circ}$C, T$_{comp}$ = -73$^{\circ}$C, and M$_{s}$ is approximately 150 emu/cc. All samples exhibited uniaxial PMA.

A subset of the films measured with magnetometry were analyzed using FEM. To measure changes in the MRO of the ferrimagnetic films, FEM experiments were carried out using scanning transmission electron microscopy (STEM) in an FEI TitanX operated at an accelerating voltage of 200 kV and a 2.2 nm diameter probe. Scanning nanodiffraction acquisition parameters are detailed in the Supplemental Materials. Additionally, the films deposited at 20$^{\circ}$C and 300$^{\circ}$C were tilted at an angle varied between 0$^{\circ}$ and 40$^{\circ}$ in 5$^{\circ}$ increments to determine whether MRO varies with orientation through the film. These two films were selected for tilting because they had the greatest difference in MRO as determined from FEM variance measurements.

FEM analysis was performed following the method outlined by Kennedy \textit{et al.}\cite{kennedy2020} Custom scripts were used to fit the radial profiles of the scanning nanodiffraction data using least squares fitting of the first diffracted ring as described by Gammer \textit{et al.} \cite{gammer2018} The fitting parameters are used to correct for any ellipticity in the patterns before calculating the variance in intensity as a function of scattering vector. For patterns collected from tilted films, the ellipticity is not corrected. Instead, changes in the major elliptical axes are used as a measure of relative changes in mean bond lengths. The procedure for collecting scanning nanodiffraction data and FEM analysis is outlined in the Supplemental Materials. For FEM, the variance is calculated with respect to scattering angle \textbf{k}, position on the sample \textit{r}, and resolution (which is a function of \textit{r}); however, in STEM FEM only the scattering vector varies and thus the equation for variance \textit{ V$_{\sigma}$} in diffracted intensity \textit{I} becomes\cite{treacy2005}:
\begin{eqnarray}
    V_{\sigma}(\textbf{k}, r) = \frac{ \langle I^2(\textbf{k},r) \rangle
    -
    \langle I(\textbf{k},r) \rangle^2}
    {\langle I(\textbf{k},r) \rangle^2},
\end{eqnarray}

Additionally, Lorentz TEM images were collected from the as-deposited film grown at 20$^{\circ}$C and the film grown at 20$^{\circ}$C and annealed at 200$^{\circ}$C and 300$^{\circ}$C to probe the effect of annealing on domain structure. Lorentz TEM was carried out in a FEI ThemIS operated at 300 kV in Lorentz mode. All films had been previously subjected to a magnetic field of approximately 2 T during FEM image acquisition, but the magnetic field of the objective lens was minimized to less that 0.1 T for Lorentz imaging. The films were tilted 15$^{\circ}$ and the beam was defocused to produce visible magnetic domains. 

\section*{\label{sec:level3}Results}

\begin{figure}[t]
\centering
\includegraphics[width=0.48\textwidth]{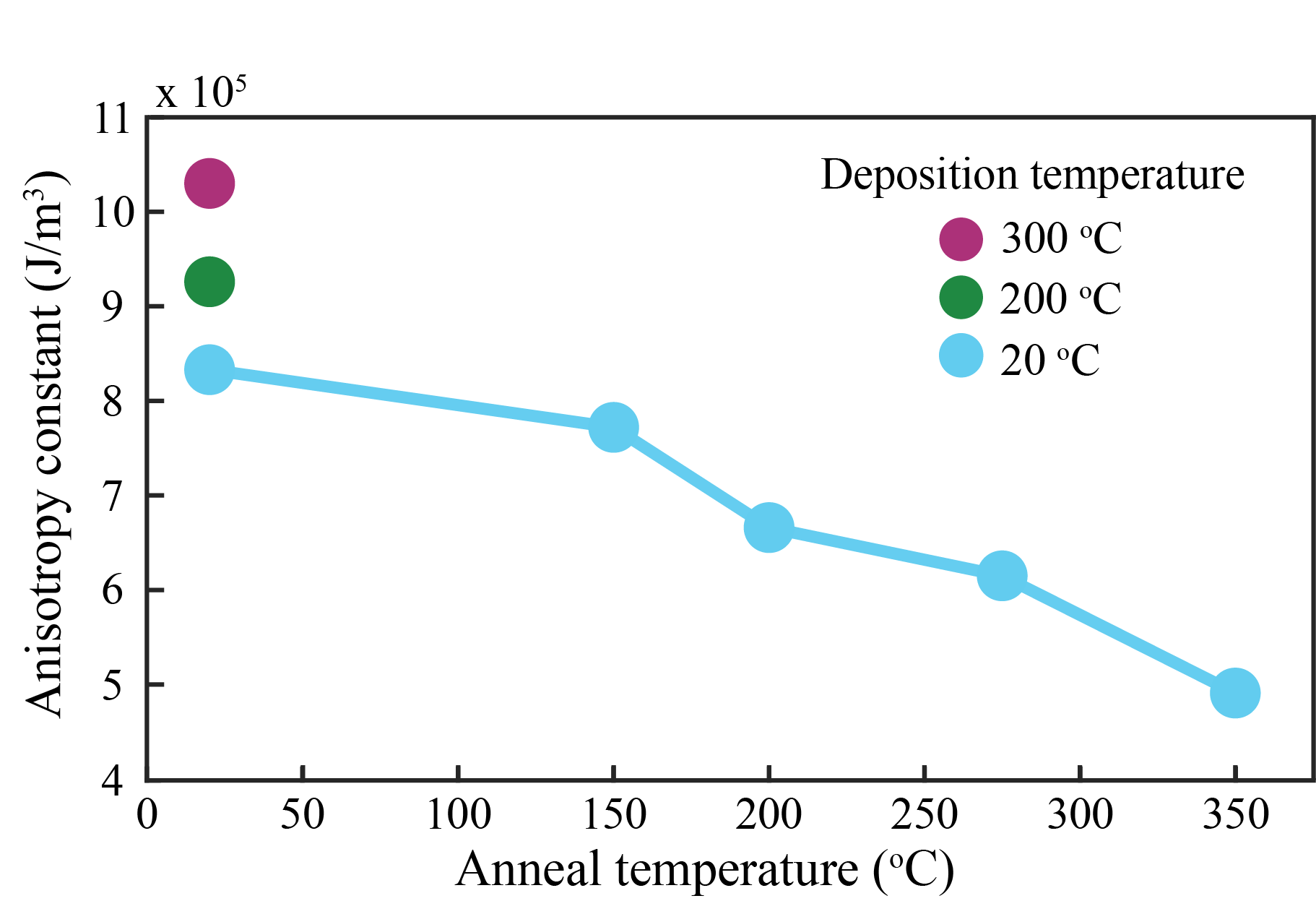}
\caption{\label{fig:fig1} Magnetic anisotropy constant as a function of deposition and annealing temperatures derived from IP and OOP magnetization measurements. M$_{s}$ is approximately the same for all films.}
\end{figure}

The magnetic anisotropy constant $K_{ui}$ is 8.33 x 10$^5$ J/m$^3$ for films grown at 20$^{\circ}$C, and increases to 9.26 x 10$^5$ J/m$^3$ for films grown at 200$^{\circ}$C and to 10.30 x 10$^5$ J/m$^3$ for films grown at 300$^{\circ}$C. For films grown at 20$^{\circ}$C and subsequently annealed at temperatures between 150$^{\circ}$C and 350$^{\circ}$C, the magnetic anisotropy decreases with temperature. The relationships between deposition temperature, annealing temperature, and magnetic anisotropy are shown in Fig.~\ref{fig:fig1}. Similar trends have been shown in other amorphous RE-TM thin film systems \cite{harris1994}. The OOP magnetic saturation $M_s$ remains constant (150 emu/cc) for all samples in the series.

High resolution TEM imaging and the mean converged beam electron diffraction (CBED) patterns confirm that the samples are amorphous (shown in Supplemental Materials). The uniformity of the diffracted speckle in the scanning nanodiffraction patterns and lack of strong Bragg scattering indicates that the films lack long range order and have no nanocrystals.

\begin{figure*}[t]
\includegraphics[width=0.80\textwidth]{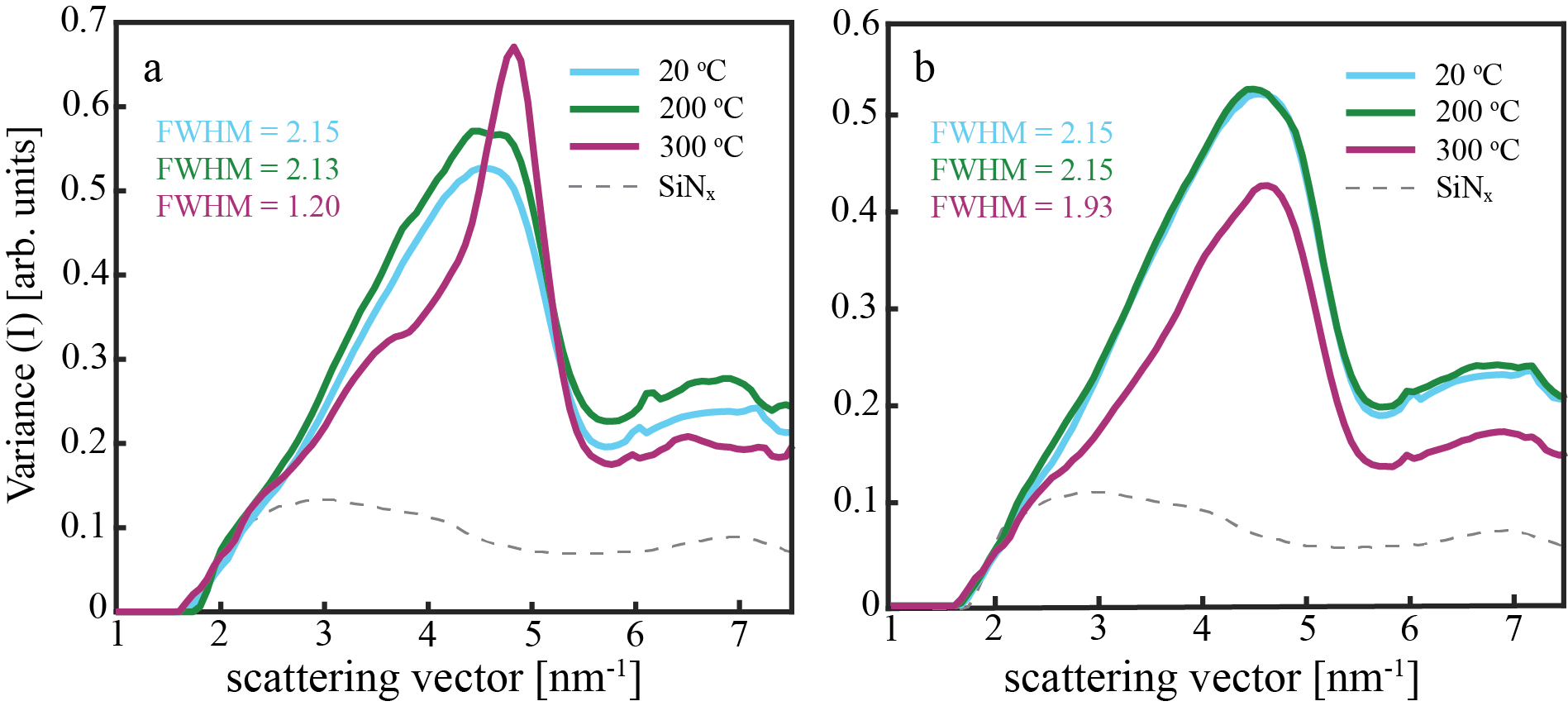}
\caption{\label{fig:fig2} Variance as a function of scattering vector \textbf{k} for \textit{a-}Tb$_{17}$Co$_{83}$ films (a) deposited at 20$^{\circ}$C, 200$^{\circ}$C, and 300$^{\circ}$C and (b) subsequently annealed at 200$^{\circ}$C and 300$^{\circ}$C following deposition at 20$^{\circ}$C. The height of the peak centered between 4.5 nm$^{-1}$ and 4.8 nm$^{-1}$ corresponds to the degree of relative MRO in the films. The magnitude of the peak increases with growth temperature, indicating that MRO is greater for films deposited at higher temperatures. The position of the peaks along the scattering vector axis indicates the average size of the bonds resulting in detectable MRO. The dashed gray lines are the variance in intensity of the SiN$_x$ control film.}
\end{figure*}

In Fig.~\ref{fig:fig2}, the normalized variance in intensity \textit{V(\textbf{k})} for films in their as-deposited state (a) and after annealing (b) is plotted as a function of scattering vector using mean statistics. The MRO increases with increasing growth temperature, as indicated by an increase in \textit{V(\textbf{k})}. The variance for the 300$^{\circ}$C growth is significantly sharper than the 20$^{\circ}$C and 200$^{\circ}$C variances.   There is an increase in the \textit{V(\textbf{k})} and a decrease in the full width at half maximum (FWHM) across the three films as a function of deposition temperature. 

\begin{table}[b]
\caption{\label{tab:table1}Summarized peak heights and positions along the scattering vector \textit{\textbf{k}} axis from FEM curves in Fig.~\ref{fig:fig2} }
\begin{ruledtabular}
\begin{tabular}{lcc}
Temperature parameters & Relative Height\footnote{The peak height of the film grown at 20$^{\circ}$C is used as the reference. Other peak heights are given relative to it.}  & Position (nm$^{-1}$)\\
\hline
20$^{\circ}$C deposition & 1 & 4.5\\
200$^{\circ}$C deposition & 1.10 & 4.5 to 4.7\\
300$^{\circ}$C deposition & 1.23 & 4.8\\
20$^{\circ}$C deposition, 200$^{\circ}$C anneal & 1.01 & 4.5\\
20$^{\circ}$C deposition, 300$^{\circ}$C anneal & 0.81 & 4.6\\
\end{tabular}
\end{ruledtabular}
\end{table}

The peak heights correspond to relative MRO and the peak positions correspond to the mean interatomic distance of the MRO. The peaks in \textit{V(\textbf{k})} are typically attributable to oriented clusters of atoms (MRO) and not to individual bond types (SRO) \cite{voyles2002,li2014}. Two main peak features emerge in the variance plots in Fig.~\ref{fig:fig2}. The first peak is centered around 2.4 nm$^{-1}$ and is attributable to \textit{a-}SiN$_x$, as confirmed through FEM analysis of 10 nm of \textit{a-}SiN$_x$ deposited onto \textit{a-}SiN$_x$ membranes. The second peak is centered between 4.5 and 3.3 nm$^{-1}$ and originates from the \textit{a-}Tb$_{17}$Co$_{83}$. The peak positions of the \textit{a-}Tb$_{17}$Co$_{83}$ growth series shift to larger \textbf{\textit{k}} values with increasing deposition temperature. The change in the magnitude of the \textit{a-}Tb$_{17}$Co$_{83}$ peak variance relates to degree of MRO in the films. Peak positions and relative heights are provided in Table~\ref{tab:table1}.

The FEM diffracted signal varies slightly as a function of orientation through the individual films, as shown in Fig.~\ref{fig:fig3}, suggesting directional dependency in bonding structure. The most significant difference in MRO as a function of tilt angle was observed between the films grown at 20$^{\circ}$C and 300$^{\circ}$C. These films were tilted in the TEM to 40$^{\circ}$C to check for directional dependency of the MRO. Changes in the ellipticity of the FEM patterns as a function of tilt angle reveal that the films deposited at 300$^{\circ}$C exhibit a greater change in the length of their major axes, compared to the major axes of the films grown at 20$^{\circ}$C. The mean bond length of both films decreases as a function of orientation away from the film normals. The decrease is more pronounced for the film deposited at 300$^{\circ}$C, indicating greater orientation-dependency of the MRO bond lengths. By assigning the changes in the FEM diffraction rings to changes in bond length in the IP and OOP directions, we can analyze the tilted  data using strain relationships (Supplemental Materials). From this, we determine that the maximum change in mean bond length is 1.43\% for the film deposited at 20$^{\circ}$C and 3.14\% for the film deposited at 300$^{\circ}$C. These values assume the trend continues past the 40$^{\circ}$C tilt angle, to the OOP orientation. This implies that the films grown at 300$^{\circ}$C have greater directional dependence of interatomic ordering.

\section*{\label{sec:level4}Discussion}

Many models have been proposed for the origin of PMA in amorphous RE-TM thin films. Postulations include the formation of columnar microstructures, anelastic strains, surface layer anisotropy, growth-induced anisotropy resulting from magnetic interactions, microlite formation, and temperature-mediated subtle alignments of atoms around the RE atoms \cite{leamy1978,dirks1979,yan1991,fu1992,clark1973,hellman1992}. The mechanism described in the latter of these models is commonly referred to as amorphous phase texturing \cite{hellman1992}. Many of these models have been disproven by varying growth parameters and measuring the resulting variations in PMA. This work supports the model of texturing in which local adatom configurations arrange themselves to minimize surface energy during the deposition process.

In \textit{a-}Tb$_{17}$Co$_{83}$, the itinerant \textit{3d}-electrons of Co influence the magnetic ordering of the localized \textit{4f}-electrons of Tb (RE) through an antiferromagnetic exchange interaction \cite{brooks1989}. The magnetic moments of the Tb \textit{4f}-electron orbitals, which are more than half-filled, align antiferromagnetically to the moments of the Co (TM) atoms. The combined effect of the resulting negative exchange together with local uniaxial anisotropy, which acts primarily on the Tb moments,  is to produce non-colinear ferrimagnetism that is Co-dominant above T$_{comp}$ and Tb-dominant below T$_{comp}$ \cite{vaskovskiy2015,finley2016}. \textit{a-}Tb$_{17}$Co$_{83}$ has a T$_{comp}$ of approximately -73$^{\circ}$C, making it Co-dominant at 20$^{\circ}$C. The Curie temperature (T$_C$) of \textit{a-}Tb$_{17}$Co$_{83}$ is approximately 190$^{\circ}$C. The PMA increases with increasing growth temperature in the \textit{a-}Tb$_{17}$Co$_{83}$ system, including for films grown above T$_C$ \cite{hansen1988}.  The same trend has been independently observed in other amorphous RE-TM systems \cite{andreenko1997,hellman1992,hajjar1990,kobayashi1983,zhang2010}. 

\begin{figure*}[t]
\includegraphics[width=0.96\textwidth]{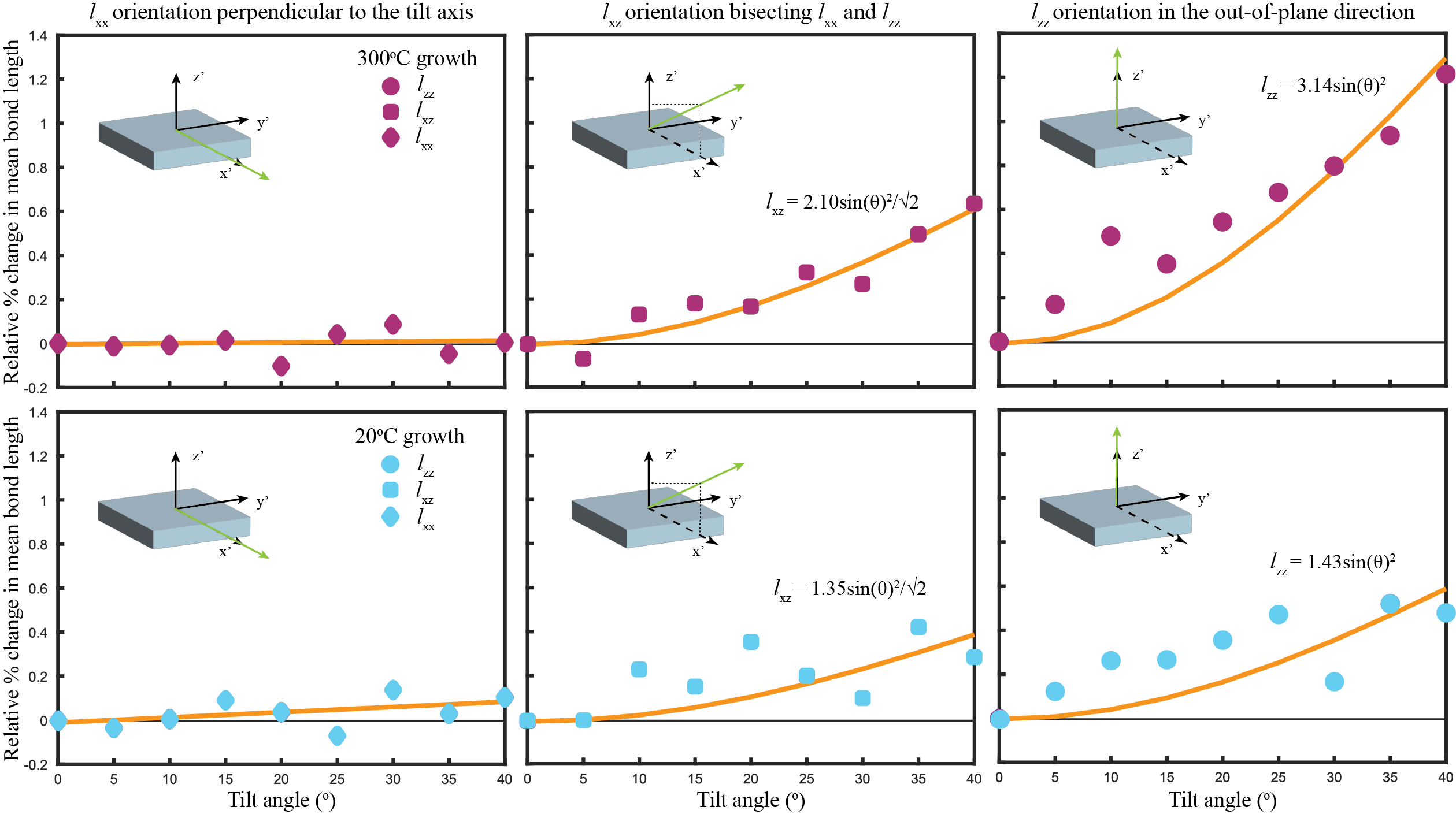}
\caption{\label{fig:fig3} Relative \% change in mean bond distances as a function of tilt angle for the films deposited at 20$^{\circ}$C and 300$^{\circ}$C. The 20$^{\circ}$C data are shown in the bottom row with blue data markers and the 300$^{\circ}$C data are shown in the top row with magenta markers. The 0$^\circ$ mean bond length is set to zero for all orientations. The z'-oriented strain changes with tilt angle as the orientation changes from IP (0$^\circ$ tilt) to OOP (extrapolated to 90$^\circ$ tilt). The film deposited at 300$^{\circ}$C exhibits a greater degree of change in relative bond lengths compared to the film deposited at 20$^{\circ}$C. The measured \% changes in mean bond length are modelled by the infinitesimal strain theory applied to a rotated system, as shown by the orange curves. The green arrows in the insets show the orientation of each \textit{l} component at 90$^{\circ}$ (experimentally inaccessible).}
\end{figure*}

There is an inverse trend between PMA and annealing temperature. Similarly, as the annealing temperature increases, the MRO decreases, as shown in Fig.~\ref{fig:fig2} and summarized in Table ~\ref{tab:table1}. Thus, increased deposition temperature increases MRO and PMA, while increased annealing temperature reduces MRO and PMA in \textit{a-}Tb$_{17}$Co$_{83}$. The \textbf{\textit{V(k)}} curves for the 20$^{\circ}$C growth and the film annealed at 200$^{\circ}$C are similar with nearly identical FWHM, position, and relative height. The similarity suggests that there is a minimum temperature required for MRO atomic rearrangement during annealing. Overcoming the energy barrier for atomic re-orientation detectable with FEM requires a long enough anneal time and high enough temperature. At 300$^{\circ}$C for one hour, the annealing parameters are sufficient to allow energy-minimizing atomic rearrangement in the bulk of the film.

Above T$_C$, the system is paramagnetic and atomic interactions are not governed by magnetic interactions. Thus, the trend in MRO for the \textit{a-}Tb$_{17}$Co$_{83}$ growth series cannot be ascribed to magnetic interactions. Instead, the relationship between MRO and growth temperature-mediated PMA supports models in which local adatom configurations arrange such that they minimize the surface energy during growth \cite{hellman1994}. As layers build, the preferential configuration is preserved, leading to oriented local magnetic anisotropy that produces an overall macroscopic PMA.

The films grown at 20$^{\circ}$C and 300$^{\circ}$C were selected for tilted FEM analysis to understand the relationship between temperature-mediated atomic texturing during deposition and orientationally anisotropic MRO \cite{kennedy2020}. In this measurement, FEM data is acquired at varying stage tilt angles, which enables us to deduce the relative change in bond lengths associated with the MRO in the IP and OOP directions. Because the diffraction patterns contain information about d-spacings perpendicular to the direction of the beam, as the effective angle between the beam and the sample is varied, any changes in bond length between the IP and OOP directions in the film will cause an ellipticity in diffraction pattern (Supplemental Materials). This effect is similar to ellipticity in diffraction data caused by anisotropic strain, although in our case, decoupling the influences of strain and changes in preferential atomic orientations is not possible. For this reason, we consider changes in the major axis of the fitted ellipse to correspond broadly to changes in the average bond length, $l$, as a function of orientation through the film. \textit{l}$_{xx}$ is along the x-axis, \textit{l}$_{zz}$ in the y-z plane along the z'-axis, and \textit{l}$_{xx}$ bisects the x' and z' directions.

 The stage tilt is limited to 40$^{\circ}$ in the TitanX, but the OOP bond length can be extrapolated from measurements at lower angles by considering the geometry of the experiment and the influence of infinitesimal shifts in bond length on the diffraction data, which gives a $\sin^2$ dependence to the projected OOP bond length. As shown in Fig.~\ref{fig:fig3}, \textit{l}$_{xx}$ is modeled with a linear fit and \textit{l}$_{xz}$ is modeled with a $\sin^2$$/\sqrt{2}$ dependence. The geometric origins of these terms are illustrated in the Supplemental Materials. 

Fig.~\ref{fig:fig3} shows the trend in average relative \% change in mean bond length as a function of tilt angle for the films grown at 20$^{\circ}$C and 300$^{\circ}$C. Relative \% change below the ~$\pm0.2\%$ margin of error indicates that the films are isotropic within the limit of the technique. Above this margin of error, the films exhibit changes in relative mean bond distances dependent on orientation through the films.

From the \textit{l}$_{zz}$ fits, the extrapolated maximum bond lengths are calculated. For the film deposited at 20$^{\circ}$C, the extrapolated maximum relative \% change in mean bond length is 1.4\%. For the film deposited at 300$^{\circ}$C, the extrapolated maximum relative \% change in mean bond length is 4.5\%. These values are determined by extending the fitting curves to a theoretical 90$^{\circ}$ tilt. Both films exhibit shorter mean bond lengths in the IP direction compared to the OOP direction. The IP bond lengths in the film grown at 300$^{\circ}$C are larger, on average, than the bonds in the OOP orientation. Assuming the texturing model for PMA, this corresponds to a greater proportion of Co-Tb bonds in the OOP orientation relative to Co-Co bonds in the IP orientations. The greater relative \% change in mean bond length for 300$^{\circ}$C indicates that at higher deposition temperatures adatoms with higher energies at impingement orient such that Co-Tb bonds form preferentially in the OOP direction, consistent with EXAFS data on \textit{a}-Tb-Fe films\cite{harris1992,harris1993}.

The dramatic decrease in PMA observed between the film grown at 20$^{\circ}$C and the films annealed at higher temperatures and the corresponding change in MRO prompted the use of Lorentz TEM. The Lorentz images, shown in Supplemental Materials, show the typical high-contrast domain patterns seen in films with strong PMA and corroborate the trends observed in variance and magnetization for \textit{a-}Tb$_{17}$Co$_{83}$ annealed at increasing temperatures \cite{hellman1999}. Higher PMA causes a high energy cost of forming domain walls and narrower domain walls, with well defined domains, while lower PMA causes wider domain walls and less defined domains. The result is sharper domain walls and better defined domains in films deposited at higher temperatures, and less defined domains in films grown at lower temperatures after annealing.

The changes in mean bond length with varying orientation through the films, the increase in MRO as the deposition temperature increases, and the subsequent decrease in MRO after annealing, collectively indicate the presence of MRO bond length anisotropy in the \textit{a-}Tb$_{17}$Co$_{83}$ films. During deposition, structural variations are introduced as a function of growth temperature. A higher deposition temperature favors a higher proportion of Tb-Co bonds in the OOP direction, which results in greater MRO and increased PMA. Annealing, on the other hand, allows for structural relaxation in the films, and atom rearrangement. Taken together, the observed trends in MRO with respect to growth conditions suggest that the evolution of PMA in the films is due to structural changes.

\section*{\label{sec:level5}Conclusions}
Numerous studies on amorphous RE-TM films have noted the relationships between growth temperature parameters and PMA, but the structural origin remains under debate. This work established a relationship between thermal growth parameters, PMA, and local atomic ordering using the TEM technique of scanning nanodiffraction. MRO increases with increasing film deposition temperature and decreases with increasing anneal temperature. Thus, there is a correlation between the degree of MRO and PMA in \textit{a-}Tb$_{17}$Co$_{83}$ films grown via magnetron co-sputtering. These results support an amorphous phase texturing model in which adatom configuration varies as a function of deposition temperature and annealing allows for subsequent relaxations of preferential configurations within the films. Tb-Co (RE-TM) pairs prefer to form vertically during deposition, but annealing allows for rearrangement of atomic pairs such that there is greater uniformity across orientations. Analysis of tilted FEM data shows a greater local ordering anisotropy in the film exhibiting the highest degree of MRO (300 $^{\circ}$C growth) compared to the the film grown at 20$^{\circ}$C with less MRO.

\begin{acknowledgments}

This work was  supported by National Science Foundation STROBE grant DMR-1548924. Work at the Molecular Foundry was supported by the Office of Science, Office of Basic Energy Sciences, of the U.S. Department of Energy under Contract No. DE-AC02-05CH11231. Growth and non-electron microscopy characterization of experimental \textit{a-}Tb$_{17}$Co$_{83}$ and \textit{a-}SiN$_x$ were performed by D.O. and E.H. and supported by the U.S. Department of Energy, Office of Science, Basic Energy Sciences, Materials Sciences and Engineering Division under Contract No. DE-AC02-05-CH11231, under the Nonequilibrium Magnetic Materials Program (KC2204). C.O. acknowledges support from the US Department of Energy Early Career Research Program.

\end{acknowledgments}

\section*{Data Availability Statement}

The data that support the findings of this study are available from the corresponding author, M.C.S., upon reasonable request. The scripts used to process the FEM data are available at https://github.com/ScottLabUCB/FEM.

\nocite{*}
\bibliography{maintext}

\end{document}


\maketitle

\pagestyle{empty}

\section*{\label{sec:level1} Supplemental figures}

\begin{figure}[!ht]%
\centering
\includegraphics[width=14 cm]{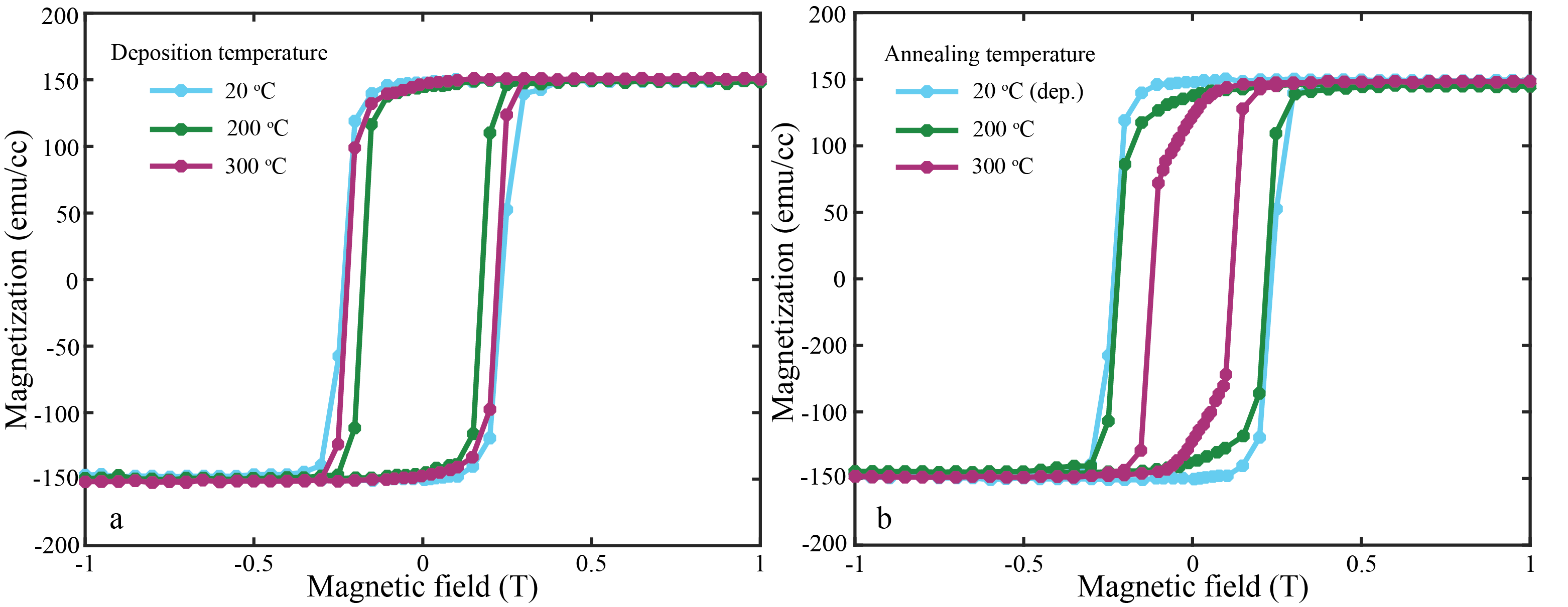}
\caption{\label{fig:fig1} (a) Magnetization M(H) with H applied out-of-plane for (a) films grown at 20$^{\circ}$C, 200$^{\circ}$C, and 300$^{\circ}$C, and (b) films grown at 20$^{\circ}$C as-deposited and after annealing at 200$^{\circ}$C and 300$^{\circ}$C. All films show 
strong perpendicular magnetic anisotropy (PMA), with constant M$_{s}$ and anisotropy constant K$_{ui}$ that increases with increasing deposition temperature  and decreases with increasing annealing temperature (as discussed in the main text and shown in Fig. 1).  As a note, the coercivity H$_c$ is significantly less (factors of ten or more) than the anisotropy field $H_k=2K_{ui}/M_s$ in all cases because it is determined by domain wall nucleation and motion, and hence H$_c$ depends on but is not directly proportional to K$_{ui}$ [Hellman 1999].}
\end{figure}

\begin{figure}[!ht]%
\centering
\includegraphics[width=0.5\textwidth]{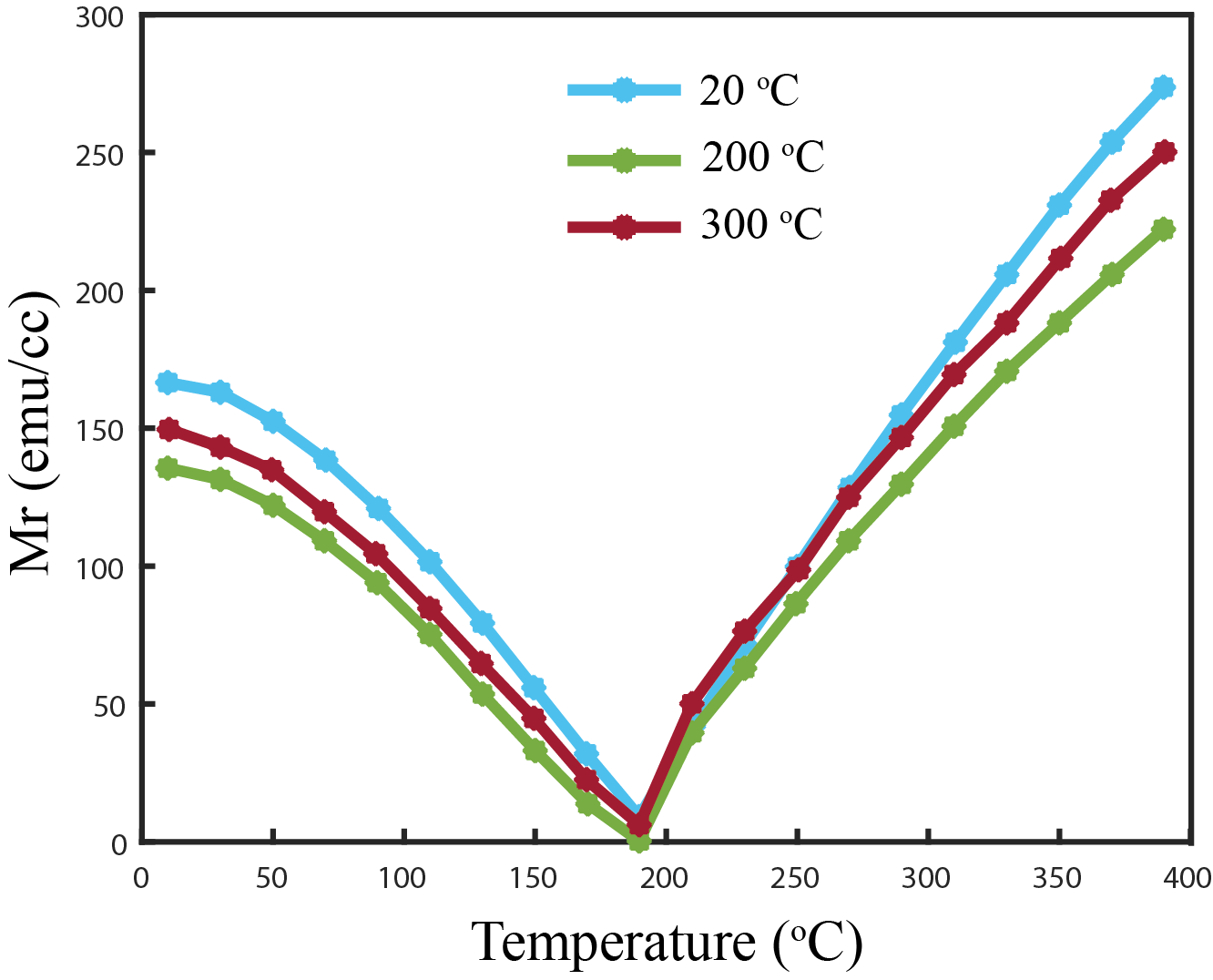}
\caption{\label{fig:fig2} Temperature dependence of the remnant magnetization for the films grown at 20$^{\circ}$C, 200$^{\circ}$C, and 300$^{\circ}$C. The films were saturated with 5 T field at each temperature point prior to measuring the remnant magnetization.}
\end{figure}

\begin{figure}[!ht]%
\centering
\includegraphics[width=0.95\textwidth]{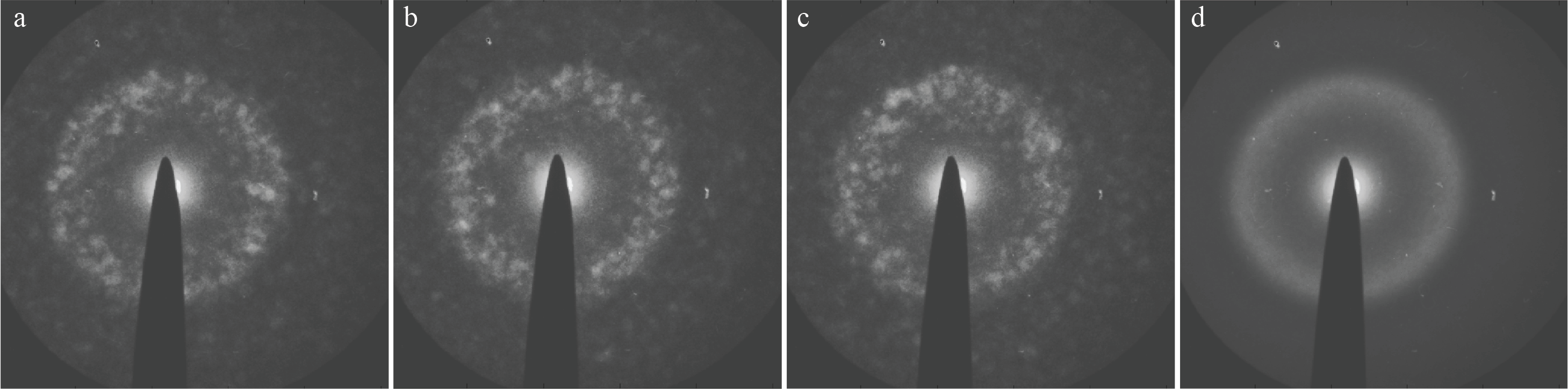}
\caption{\label{fig:fig3} (a-c) Sample scanning nanodiffraction images displaying the characteristic speckle pattern of amorphous materials. These images were taken from the \textit{a-}Tb$_{17}$Co$_{83}$ film deposited at room temperature. The electron probe is positioned such that the ring patterns do not overlap defects on the detector. A beam stop covers the central beam and the beam remains in the same location in all FEM images. (d) Mean scanning nanodiffraction image produced by averaging all images in a stack of 144 images. The mean pattern is used for calculating V(\textbf{\textit{k}}).}
\end{figure}

\begin{figure}[!ht]%
\centering
\includegraphics[width=0.98\textwidth]{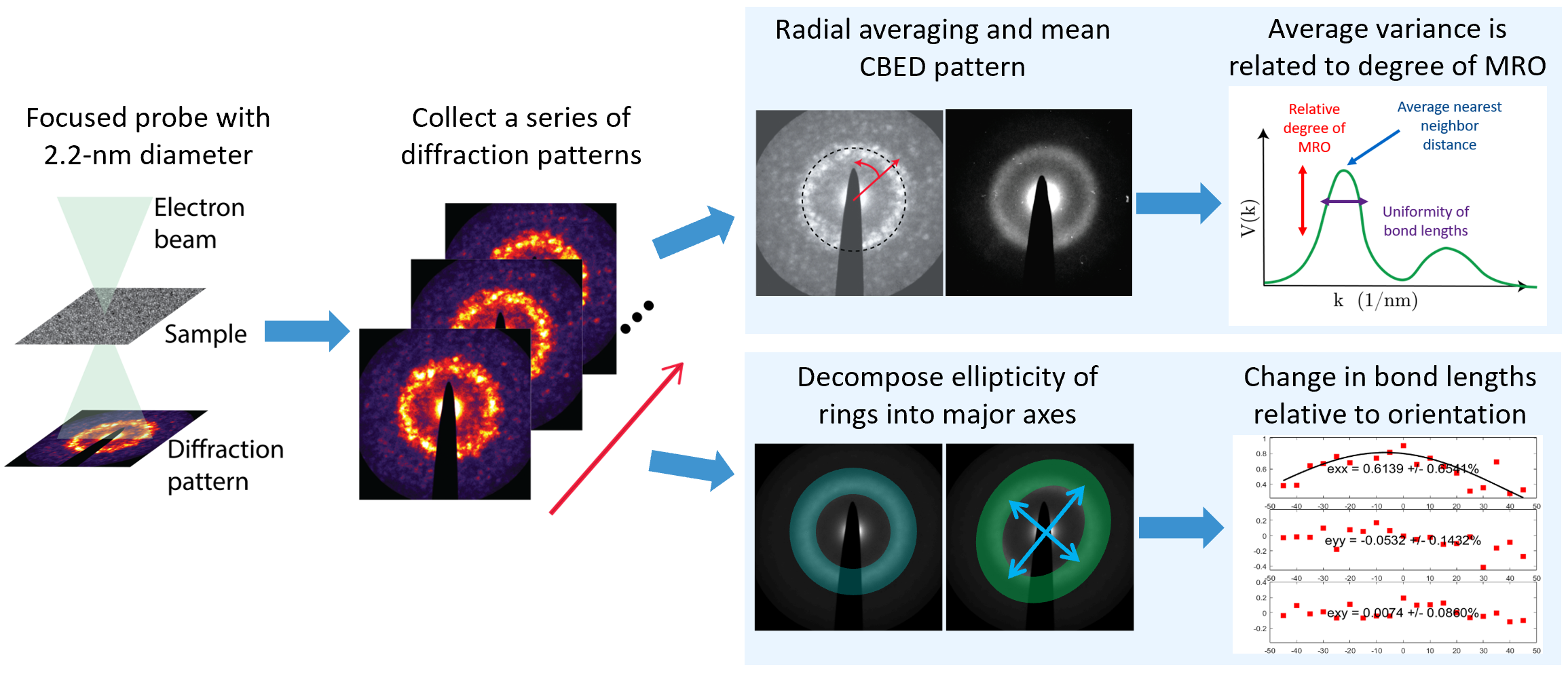}
\caption{\label{fig:fig4} Workflow for collecting scanning nanodiffraction data and FEM analysis of variance in speckled intensity. The electron beam is focused to a probe in STEM mode and the sample is placed at eucentric height to produce the characteristic speckled pattern of FEM. The beam is rastered across multiple regions of the film to collected a series of scanning nanodiffraction patterns. }
\end{figure}

\begin{figure}[!ht]%
\centering
\includegraphics[width=0.9\textwidth]{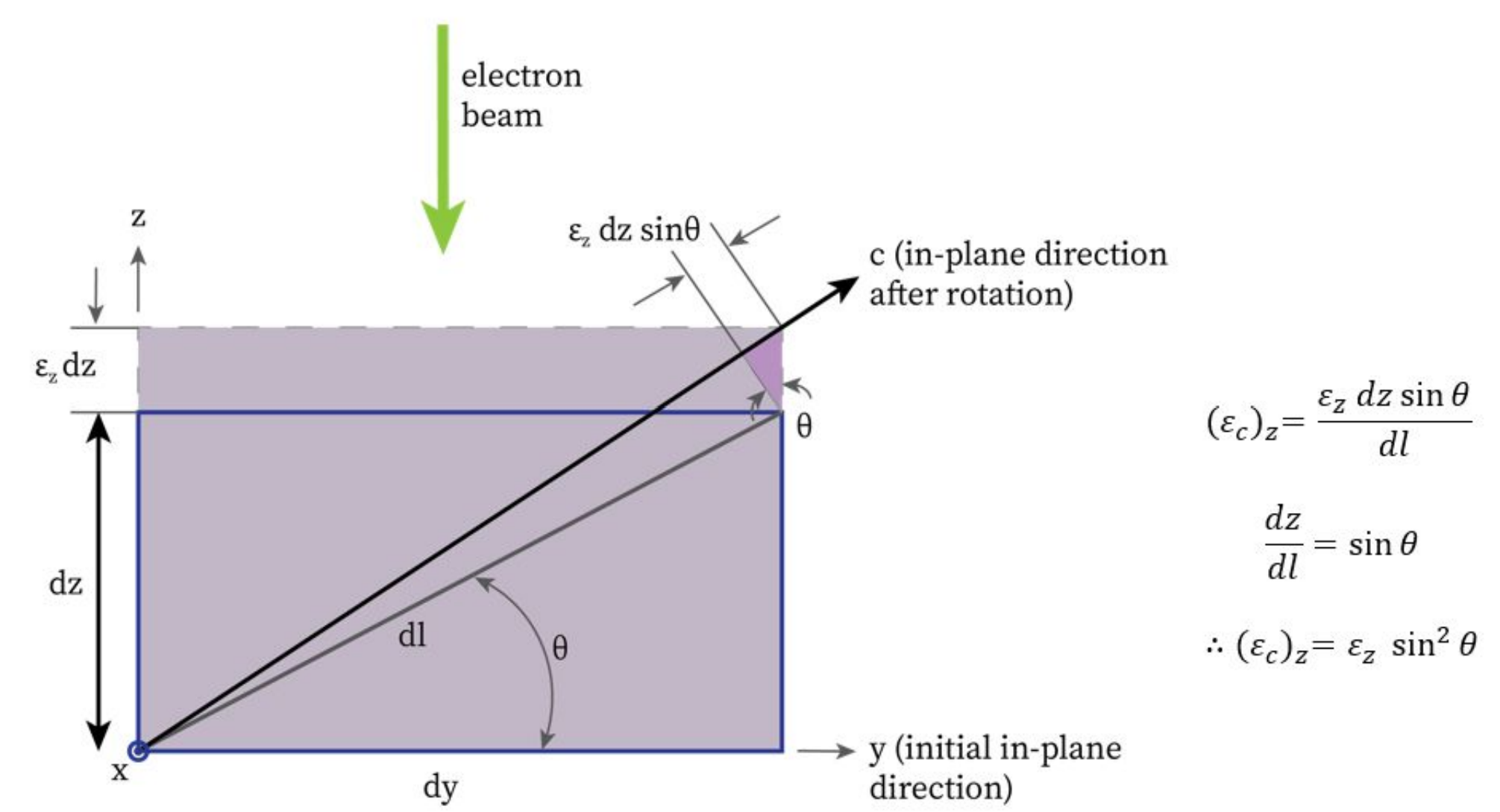}
\caption{\label{fig:fig5}The geometry of the tilted FEM experiment and origin of the $sin^2(\theta)$ dependence on tilt angle of the out of plane bond length in the film. }
\end{figure}

\begin{figure}[b]%
\centering
\includegraphics[width=0.95\textwidth]{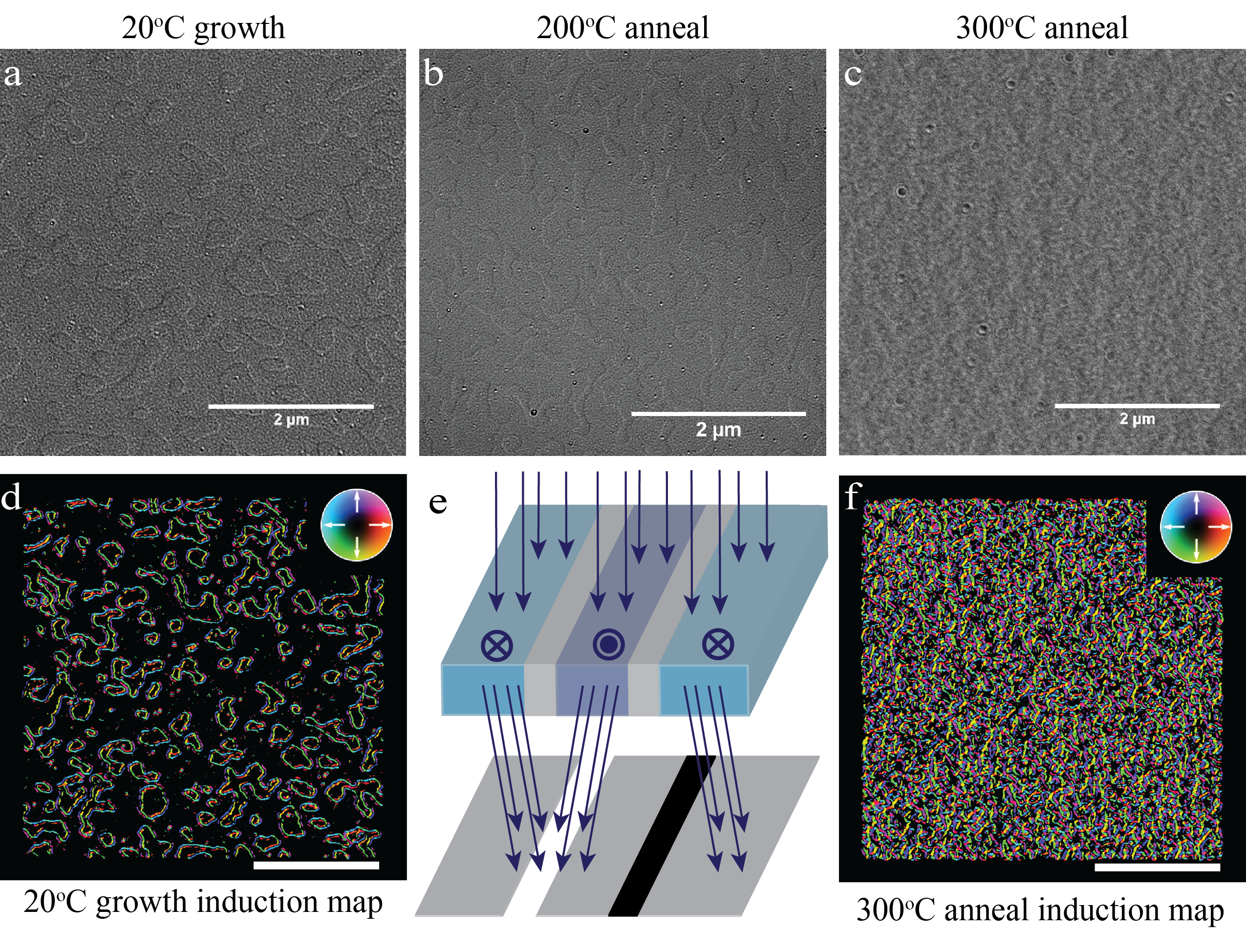}
\caption{\label{fig:fig6}(a-c) Lorentz TEM images collected from the films (a) deposited at room temperature, (b) annealed at 200$^{\circ}$C, and (c) annealed at 300$^{\circ}$C. Induction maps corresponding to (a) and (c) are shown in (d) and (f), respectively. The domains are significantly denser and less defined in the film annealed at 300$^{\circ}$C compared to the films deposited at room temperature and annealed at 200$^{\circ}$C. A larger defocus was required to to view the domain structure in the film annealed at 300$^{\circ}$C compared to the two other films depicted. (e) A schematic depiction of the electron beam interacting with a magnetic sample and deflecting to produce contrast in Lorentz TEM. All scale bars are 2 $\mu$m.}
\end{figure}

\clearpage

\section*{\label{sec:level2} Fluctuation Electron Microscopy Workflows}
\subsection*{\label{sec:level3} Overview of Fluctuation Electron Microscopy}

FEM is based on scanning electron nanodiffraction measurements, shown in Supplemental Materials Figures~\ref{fig:fig3} and \ref{fig:fig4}. There are two pathways for analysis following the collection of a series of scanning nanodiffraction patterns. The first route is to correct the ellipticiy of the patterns and calculate the variance in intensity as a function of scattering vector. This route yields a  measure of the relative amount of MRO in the samples.  The second route is to measure relative changes in the major axes of the fitting ellipses for a series of nanodiffraction pattern stacks collected across samples of varied conditions or orientations to obtain information on relative atomic spacings. This approach is used to determine relative changes in bond length in the in- and out-of-plane directions of the film.

\subsection*{\label{sec:level5} Acquiring scanning nanodiffraction data}

Experiments were carried out in a TitanX operated at 200 kV. The third condenser lens was set to produce a convergence angle of 0.51 mrad resulting in a measured 2.2 nm diameter probe and a 15.5 pA probe current. Images were collected on an Orius charge coupled device (CCD) detector with an exposure time of 0.3 seconds and a camera length of 300 mm. The images were binned by a factor of four, to a final size of 512x512 pixels. Nanodiffraction data were collected as 12x12 image stacks (144 images per stack) for each film with 5 nm step sizes between regions of analysis. Collection was repeated four times per film (576 images in total per film) for statistical averaging  over different regions of the film and to avoid excess contamination. During the experiment, all diffraction patterns were taken with the same imaging conditions to ensure that the effects of microscope misalignment were the same across all patterns. The small feature that appears around 4 nm$^{-1}$ in all variance curves is from a defect on the CCD. Additionally, the films deposited at 20$^{\circ}$C and 300$^{\circ}$C were tilted at an angle varied between 0$^{\circ}$ and 40$^{\circ}$ in 5$^{\circ}$ increments to determine whether MRO varies with orientation through the film. These two films were selected for tilting because they had the greatest difference in MRO as determined from FEM variance measurements.

\subsection*{\label{sec:level5} Measurements Using Tilted Fluctuation Electron Microscopy }

The geometry of the tilted FEM experiment are shown in Supplemental Materials Fig.~\ref{fig:fig5}. The electron beam is incident along the z direction, which is identical to the out-of-plane direction of the untilted film. The x direction (out of the page) is the tilt axis, and the y direction in the diagram is perpendicular to the tilt axis. Since the beam samples bond lengths in the directions perpendicular to the propagation direction, changing the sample tilt angle samples different directions in the film. Specifically, tilting the sample probes bond lengths that lie out of plane in the direction perpendicular to the tilt axis. Another explanation is that as the film is tilted with respect to the beam, the Ewald sphere samples increasingly out-of-plane bond lengths in one direction. The result is that if there are differences in bond length between the in and out of plane directions, then a previously round diffraction pattern would become an ellipse.  From the fitted ellipse, the relative difference between the in- and out-of-plane bond-length can be determined. 

To determine this difference in bond lengths, we model the  measured strain using infinitesimal strain theory applied to a rotated system. For \textit{a}-Tb-Co, we refer to \% changes in mean bond length (\textit{l}$_{zz}$) instead of strain ($\epsilon_{zz}$) because decoupling the influences of strain and changes in preferential atomic orientations is not possible. The relative \% change in bond length in the y-z plane along the z'-direction is denoted \textit{l}$_{zz}$ and relative \% change in bond length in the x'-direction along the x-axis is \textit{l}$_{xx}$. The x'-direction is the in-plane direction that remains constant while the z'-direction is the out-of-plane direction used to determine relative changes in bond lengths. The z'-direction is within the plane of the film when untilted and perpendicular to the film at a tilt of 90$^{\circ}$ (experimentally inaccessible).

If  $\epsilon_z dz$ is the infinitesimal displacement due to an effective strain epsilon, the \% change in bond length in the ``c'' direction $dl$ after a rotation $\theta$ is equal to $\epsilon_z \cdot dz \sin^2(\theta)$. Therefore, the effective strain in the c direction $\frac{\epsilon_z \cdot dz \sin^2(\theta)}{dl}$. From geometry, $\frac{dz}{dl}$ is equal to $\sin(\theta)$, therefore the effective strain in the c direction due to $\epsilon_z dz$, $(\epsilon_c)_z$ is  $\epsilon_z \sin(\theta) $. Although our approach is based on continuous strain theory, we replace the variable $\epsilon$ with $l$ to indicate that our measurement is more general, and captures the changes in average bond length regardless of their origin. 

Based on this dependence on $\theta$, we took several FEM measurements at different tilt angles. The observed changes in the major axis of the elliptical fits of the diffraction rings give experimental observations of the relative changes in mean bond length in the direction perpendicular to the electron beam. The extapolated value of $l_{zz}$ is obtained by fitting according to the $\sin^2(\theta)$ dependence derived above. 

\section*{\label{sec:level5} Supplemental Materials References}
F. Hellman, A. L. Shapiro, E. N. Abarra, P. W. Rooney, and M. Q. Tran, “Magnetic Anisotropy and Coercivity in Magneto-Optical Recording Materials,” Journal of the Magnetics Society of Japan \textbf{23} 79-84 (1999).
